%% file: main.tex
\documentclass[journal]{IEEEtran}
\input{packages}
\input{macros}

\input{abbreviations}

\begin{document}

\title{Learning Radio Environments by\\ Differentiable Ray Tracing}

\author{
    Jakob Hoydis,    
    Fay\c{c}al A\"it Aoudia,
    Sebastian Cammerer,
    Florian Euchner,
    Merlin Nimier-David,\\
    Stephan ten Brink, and
    Alexander Keller

    \thanks{J.\ Hoydis and F.\ A\"it Aoudia are with NVIDIA, France (email: \{jhoydis, faitaoudia\}@nvidia.com).}
    
    \thanks{S.\ Cammerer and A. Keller are with NVIDIA, Germany (email: \{scammerer, akeller\}@nvidia.com).}
    
    \thanks{M.\ Nimier-David is with NVIDIA, Switzerland (email: mnimierdavid@nvidia.com).}
    
    \thanks{F.\ Euchner and S.\ ten Brink are with the University of Stuttgart, Germany, (email: \{euchner, tenbrink\}@inue.uni-stuttgart.de) and acknowledge support by the German Federal Ministry of Education and Research (BMBF) within the project KOMSENS-6G (grant no. 16KISK113).}

    \thanks{J.\ Hoydis, F.\ A\"it Aoudia, and S.\ Cammerer acknowledge financial support from the European Union under Grant Agreement 101096379 (CENTRIC).}
}
\maketitle
\input{sections/abstract}
\glsresetall
\input{sections/introduction}

\input{sections/rt}

\input{sections/diff-rt}
\input{sections/experiments}
\input{sections/conclusions}

\bibliographystyle{IEEEtran}
\bibliography{IEEEabrv, bibliography}
\end{document}

%% file: packages.tex
\usepackage{amsmath,amssymb, mathtools, bm}
\usepackage{url}
\usepackage[hidelinks]{hyperref}
\usepackage{graphicx}
\usepackage{subcaption}
\usepackage{glossaries}
\usepackage{siunitx}
\usepackage[ruled,boxed,lined,linesnumbered]{algorithm2e}
\usepackage{flushend}

%% file: macros.tex
\renewcommand{\vec}[1]{{\mathbf{#1}}}
\newcommand{\vecs}[1]{{\boldsymbol{#1}}}

\newcommand{\bv}{\vec{b}}

\newcommand{\ev}{\vec{e}}

\newcommand{\hv}{\vec{h}}

\newcommand{\kv}{\vec{k}}

\newcommand{\nv}{\vec{n}}

\newcommand{\pv}{\vec{p}}

\newcommand{\rv}{\vec{r}}
\newcommand{\sv}{\vec{s}}

\newcommand{\vv}{\vec{v}}
\newcommand{\wv}{\vec{w}}
\newcommand{\xv}{\vec{x}}
\newcommand{\yv}{\vec{y}}
\newcommand{\zv}{\vec{z}}

\newcommand{\muv}{\vecs{\mu}}

\newcommand{\varphiv}{\vecs{\varphi}}

\newcommand{\thetav}{\vecs{\theta}}

\newcommand{\Cm}{\vec{C}}
\newcommand{\Dm}{\vec{D}}
\newcommand{\Em}{\vec{E}}
\newcommand{\Fm}{\vec{F}}

\newcommand{\Qm}{\vec{Q}}

\newcommand{\Tm}{\vec{T}}

\newcommand{\Omegam}{\vecs{\Omega}}

\newcommand{\Bc}{{\cal B}}

\newcommand{\Dc}{{\cal D}}

\newcommand{\Lc}{{\cal L}}

\newcommand{\CC}{\mathbb{C}}

\newcommand{\RR}{\mathbb{R}}

\newcommand{\htp}{^{\mathsf{H}}}
\newcommand{\tp}{^{\mathsf{T}}}

\newcommand{\LB}{\left(}
\newcommand{\RB}{\right)}

\newcommand{\LSB}{\left[}
\newcommand{\RSB}{\right]}

\newcommand{\sigmoid}{\mathop{\mathrm{sigmoid}}}

%% file: abbreviations.tex
\newacronym{AABB}{AABB}{axis-aligned bounding box}
\newacronym{ACM}{ACM}{adaptive coding and modulation}
\newacronym{ADC}{ADC}{analog-to-digital conversion}
\newacronym{AGC}{AGC}{automatic gain control}
\newglossaryentry{AoA}
{
  name={AoA},
  description={angle of arrival},
  first={\glsentrydesc{AoA} (\glsentrytext{AoA})},
  firstplural={angles of arrival (\glsentrytext{AoA})},
  plural={AoA}
}
\newglossaryentry{AoD}
{
  name={AoD},
  description={angle of departure},
  first={\glsentrydesc{AoD} (\glsentrytext{AoD})},
  firstplural={angles of departure (\glsentrytext{AoD})},
  plural={AoD}
}
\newacronym{API}{API}{application programming interface}
\newacronym{AWGN}{AWGN}{additive white Gaussian noise}
\newacronym{AI}{AI}{artificial intelligence}
\newacronym{BER}{BER}{bit error rate}
\newacronym{BEC}{BEC}{binary erasure channel}
\newacronym{BLER}{BLER}{block error rate}
\newacronym{BP}{BP}{backpropagation}
\newacronym{BPA}{BP}{belief propagation}
\newacronym{BPTT}{BPTT}{back-propagation through time}
\newacronym{CDF}{CDF}{cummulative distribution function}
\newacronym{CFO}{CFO}{carrier frequency offset}
\newacronym{CFR}{CFR}{channel frequency response}
\newacronym{CN}{CN}{check node}
\newacronym{CIR}{CIR}{channel impulse response}
\newacronym{CNN}{CNN}{convolutional neural network}
\newacronym{CND}{CND}{check node decoder}
\newacronym{ConvNet}{ConvNet}{convolutional neural network}
\newacronym{CP}{CP}{cyclic prefix}
\newacronym{CRC}{CRC}{cyclic redundancy check}
\newacronym{CSI}{CSI}{channel state information}
\newacronym{DAC}{DAC}{digital-to-analog conversion}
\newacronym{DL}{DL}{deep learning}
\newacronym{DFT}{DFT}{discrete Fourier transform}
\newacronym{DOA}{DOA}{direction of arrival}
\newacronym{EMA}{EMA}{exponentially moving average}
\newacronym{FFT}{FFT}{fast Fourier transform}
\newacronym{FEC}{FEC}{forward error correction}
\newacronym{GAN}{GAN}{generative adversarial network}
\newacronym{GCS}{GCS}{global coordinate system}
\newacronym{GRU}{GRU}{gated recurrent unit}
\newacronym{GPU}{GPU}{graphic processing unit}
\newacronym{GPS}{GPS}{global positioning system}
\newacronym{CPU}{CPU}{central processing unit}
\newacronym{HDPC}{HDPC}{high-density parity-check}
\newacronym{iid}{i.i.d.\@}{independent and identically distributed}
\newacronym{IFFT}{IFFT}{inverse fast Fourier transform}
\newacronym{IDFT}{IDFT}{inverse discrete Fourier transform}
\newacronym{ISAC}{ISAC}{integrated sensing and communications}
\newacronym{KL}{KL}{Kullback-Leibler}
\newacronym{LDPC}{LDPC}{low-density parity-check}
\newacronym{LLR}{LLR}{log-likelihood ratio}
\newacronym{LOS}{LoS}{line-of-sight}
\newacronym{LS}{LS}{least squares}
\newacronym{LSTM}{LSTM}{long short-term memory}
\newacronym{LMMSE}{LMMSE}{linear minimum mean squared error}
\newacronym{MIMO}{MIMO}{multiple-input multiple-output}
\newacronym{ML}{ML}{machine learning}
\newacronym{MLE}{MLE}{maximum likelihood estimation}
\newacronym{MLP}{MLP}{multilayer perceptron}
\newacronym{MRC}{MRC}{maximum ratio combining}
\newacronym{MSE}{MSE}{mean squared error}
\newacronym{MPA}{MPA}{message passing algorithm}
\newacronym{MMSE}{MMSE}{minimum mean squared error}
\newacronym{NLP}{NLP}{natural language processing}
\newacronym{NN}{NN}{neural network}
\newacronym{NLOS}{NLoS}{non-line of sight}
\newacronym{NeRF}{NeRF}{neural radiance field}
\newacronym{OFDM}{OFDM}{orthogonal frequency-division multiplexing}
\newacronym{OSS}{OSS}{open-source software}
\newacronym{pdf}{pdf}{probability density function}
\newacronym{pmf}{pmf}{probability mass function}
\newacronym{ReLU}{ReLU}{rectified linear unit}
\newglossaryentry{RIS}
{
  name={RIS},
  description={reconfigurable intelligent surface},
  first={\glsentrydesc{RIS} (\glsentrytext{RIS})},
  firstplural={\glsentrydesc{RIS}s (\glsentrytext{RIS})},
  plural={RIS}
}
\newacronym{RL}{RL}{reinforcement learning}
\newacronym{RMS}{RMS}{root mean square}
\newacronym{RNN}{RNN}{recurrent neural network}
\newacronym{RT}{RT}{ray tracing}
\newacronym{RTN}{RTN}{radio transformer network}
\newacronym{SDR}{SDR}{software defined radio}
\newacronym{SER}{SER}{symbol error rate}
\newacronym{SFO}{SFO}{sampling frequency offset}
\newacronym{SMAPE}{SMAPE}{symmetric mean absolute percentage error}
\newacronym{SNR}{SNR}{signal-to-noise ratio}
\newacronym{SIR}{SIR}{signal-to-interference ratio}
\newacronym{SINR}{SINR}{signal-to-interference-plus-noise ratio}
\newacronym{SGD}{SGD}{stochastic gradient descent}
\newacronym{SPA}{SPA}{sum product algorithm}
\newacronym{VN}{VN}{variable node}
\newacronym{VND}{VND}{variable node decoder}
\newacronym{SVM}{SVM}{support vector machine}
\newacronym{TDOA}{TDOA}{time difference of arrival}
\newacronym{TOA}{TOA}{time of arrival}
\newacronym{UAV}{UAV}{unmanned aerial vehicle}
\newacronym{UTD}{UTD}{uniform theory of diffraction}
\newacronym{UWB}{UWB}{ultra-wideband}
\newacronym{wrt}{w.r.t.\@}{with respect to}
\newacronym{XPD}{XPD}{cross polarization discrimination}
\newacronym{ZF}{ZF}{zero forcing}

%% file: sections/abstract.tex
\begin{abstract}
    \Gls{RT} is instrumental in 6G research in order to generate spatially-consistent and environment-specific \glspl{CIR}. While acquiring accurate scene geometries is now relatively straightforward, determining material characteristics requires precise calibration using channel measurements. We therefore introduce a novel gradient-based calibration method, complemented by differentiable parametrizations of material properties, scattering and antenna patterns. Our method seamlessly integrates with differentiable ray tracers that enable the computation of derivatives of \glspl{CIR} with respect to these parameters. Essentially, we approach field computation as a large computational graph wherein parameters are trainable akin to weights of a \gls{NN}. We have validated our method using both synthetic data and real-world indoor channel measurements, employing a distributed \gls{MIMO} channel sounder. 
\end{abstract}

%% file: sections/introduction.tex
\section{Introduction}

\Gls{RT} is a mature technology capable of accurately predicting the propagation of electromagnetic waves in many scenarios of interest \cite{iskander2015}. In contrast to the widely used geometry-based stochastic channel models, e.g., \cite{3gpp38901}, \gls{RT} is able to generate spatially consistent \glspl{CIR} for specific radio environments. This is a key requirement for many 6G research topics, such as \gls{ISAC} \cite{sensing}, multi-modal sensing \cite{alkhateeb2022deepsense}, \glspl{RIS} \cite{ris}, radio-based localization \cite{studer2018channel}, \gls{ML}-based transceiver algorithms \cite{aiai}, as well as some use-cases of the 3GPP study-item on \gls{ML} for the air interface \cite{sidai2021}.  \Gls{RT} is also considered an important building block of digital twin networks \cite{alkhateeb2023real, lin20226g}. 

Research on \gls{RT} has seen a surge in interest due to \gls{ML}'s potential to speed up and improve the accuracy of calculations \cite{deepray, ray-launching-wave-prop}. Several studies have applied  \glspl{NN} for improved path loss prediction \cite{zhang2020cellular, yapar-u-net, 9954403, 9722715}, while \cite{winert} explores \glspl{NN} for modeling how rays interact with surfaces. The influence of \gls{ML} and \glspl{NN} is also reshaping the field of computer vision, notably through neural rendering techniques such as \glspl{NeRF} \cite{mueller2022instant}. Simultaneously, the interest in inverse rendering and differentiable ray tracing methods grows \cite{diffrt2018, Jakob2020DrJit, Mitsuba3}.

The accuracy of RT predictions is fundamentally reliant on two key elements: the scene geometry and the characteristics of the materials involved, e.g., their conductivity $\sigma$ and relative permittivity $\varepsilon_r$. While scene geometries can nowadays be rather easily obtained with high precision \cite{neuralangelo}, there is no automated and scalable solution to obtain material characteristics. Finding these is a process know as \emph{calibration}.

There are essentially three different approaches to ray tracer calibration in the literature.
The first consists in measuring the material characteristics of individual objects \cite{8436011} of which a scene is composed. While this approach is the only option when the acquisition of channel measurements is difficult, it does not scale to large and complex scenes. It is also independent of the actual \gls{RT} process and hence cannot compensate for any mismatches in the geometry or field computations.

The second approach tries to tune material parameters to fit aggregate statistics of rays, such as path loss or delay spread, to measurements. In \cite{kanhere2023calibration}, a simplified \gls{RT} model based on angle-independent reflection and penetration losses is proposed which allows for a closed-form calibration solution. The authors of \cite{jemai2009calibration} use simulated annealing to optimize the relative permittivity and loss tangent for an \gls{UWB} system in an indoor scenario. The idea has been extended to account for diffuse reflections in \cite{6884087} and used in \cite{8168415} for the calibration of a high-speed railway scenario. 

The third approach aims at extracting individual rays from measurements which can then be matched to the \gls{RT} predictions. In \cite{charbonnier2020calibration}, a sophisticated calibration effort of a commercial ray tracer against outdoor high-precision measurements at \SI{28}{\giga\hertz} is described. The novelty \gls{wrt} prior work is the calibration on a per-ray basis rather than on aggregate statistics which allows for the object-specific calibration of diffuse scattering parameters. A key step in the process is the extraction of rays from measurements in the angle-delay domain using super-resolution techniques and matching them to the \gls{RT}-based predictions. The same technique was applied in \cite{10266601} to an indoor industrial scenario.

The aim of this paper is to introduce a novel fourth calibration method that leverages the automatic gradient computation capabilities of differentiable ray tracers like Sionna~RT \cite{sionna-rt}. After a suitable parametrization of the material characteristics, antenna and scattering patterns, the field computation can be seen as a large computational graph, similar to an \gls{NN}, with measurement positions as inputs and the corresponding \glspl{CFR} as outputs. The trainable parameters of this graph can then be optimized \gls{wrt} any differentiable loss function of the outputs using the same gradient-based methods which have led to the success of deep \glspl{NN}, e.g., \cite{adam}. However, in contrast to \glspl{NN}, the calibrated model is fully explainable as each parameter has a precise physical meaning. One may therefore think of our calibration approach as a special kind of physics-informed \cite{raissi2019physics} or model-based \gls{NN}. It scales to very complex scenes and allows for learning of antenna and scattering patterns which is out of reach for existing methods. The dataset, 3D model of the scene, and source code to reproduce all results presented in this paper are provided in \cite{diff-rt-calibration} and \cite{dichasus_dataset}.

The rest of this paper is structured as follows. Section~\ref{sec:background} discusses relevant background on \gls{RT}. Section~\ref{sec:diff_rt} describes different parametrizations of materials, antenna and scattering patterns, and introduces the loss function used for training. Section~\ref{sec:experiments} contains a description of experiments with synthetic and measured data. Section~\ref{sec:conclusions} concludes the paper.

\textit{Notations:} Vectors and matrices are denoted by boldface letters. Unit vectors are represented by a hat symbol, e.g., $\hat{\xv}$.
In a \gls{GCS} with Cartesian standard basis $(\hat{\xv},\hat{\yv},\hat{\zv})$, we denote by $\hat{\rv}(\theta, \varphi)$, $\hat{\thetav}(\theta, \varphi)$, and $\hat{\varphiv}(\theta, \varphi)$ the spherical unit vectors.

%% file: sections/rt.tex
\section{Ray tracing background}\label{sec:background}
This section explains how a \gls{CIR} is obtained in the open-source differentiable ray tracer Sionna RT \cite{sionna-rt}. While most of this material can be found in well-known text books and tutorial papers, we have decided to reproduce it here so that the reader can clearly understand which parameters can be calibrated, how they impact the \gls{CIR}, and why the latter is differentiable \gls{wrt} to them.

\subsection{Channel impulse response (CIR)}
The \gls{CFR} $H(f)$ between a transmitter and receiver at carrier frequency $f$ is computed as the sum of $M$ propagation paths (e.g., \cite[Eq. (8)]{fugen2006capability}):
\begin{align}\label{eq:cfr_cont}
    \hspace{-5pt}H(f) = \sum_{i=1}^M \underbrace{\frac{\lambda}{4\pi}\Cm_\text{R}\LB \theta^{\text{R}}_{i}, \varphi^\text{R}_{i}\RB\htp \Tm_i\LB \Cm_\text{T}\LB \theta^\text{T}_i, \varphi^\text{T}_i\RB\RB}_{\triangleq a_i} e^{-\mathrm{j}2\pi f \tau_i}
\end{align}
where $\lambda$ is the wavelength, $(\theta^\text{T}_i, \varphi^\text{T}_i)$ and $(\theta^{\text{R}}_{i}, \varphi^\text{R}_{i})$  are the \glspl{AoD} and \glspl{AoA} of the $i$th path with delay $\tau_i$ and transfer function $\Tm_i: \CC^{2}\mapsto \CC^2$. $\Cm_\text{T}(\theta,\varphi)\in\CC^2$ and $\Cm_\text{R}(\theta,\varphi)\in\CC^2$ are the transmit- and receive-antenna patterns, respectively. All of these terms will be explained in detail later on. Each path can hence be described by a complex coefficient $a_i\in\CC$ and delay $\tau_i$.

For an ideal \gls{OFDM} system with $N$ (even) subcarriers spread $\Delta_f$ apart, the channel is characterized by the discrete \gls{CFR} $H[n]$:
\begin{align}
H[n] = H(f+n \Delta_f),\, n=-\frac{N}{2},\dots,\frac{N}{2}-1.
\end{align}
We denote by $W=N\Delta_f$ the bandwidth. The discrete complex baseband-equivalent \gls{CIR} $h[\ell]$ is obtained from $H[n]$ via an $N$-point \gls{IDFT}, i.e.,
\begin{align}
    h[\ell] = \frac{1}{\sqrt{N}}\sum_{n=-\frac{N}{2}}^{\frac{N}{2}-1} H[n] e^{\mathrm{j}\frac{2\pi}{N}n\ell},\, \ell=-\frac{N}{2},\dots,\frac{N}{2}-1.
\end{align}
The vector with elements $h[\ell]$ is denoted ${\hv\in\CC^N}$.

\subsection{Modeling of antennas}\label{sec:antenna_modeling}
Antennas are described by their complex antenna pattern $\Cm(\theta, \varphi)\in\CC^2$ which consists of a vertical $\hat{\thetav}(\theta, \varphi)$ and horizontal $\hat{\varphiv}(\theta, \varphi)$ component:
\begin{align}
    \Cm(\theta, \varphi) = \begin{bmatrix} C_\theta(\theta, \varphi) & C_\varphi(\theta, \varphi) \end{bmatrix}\tp.
\end{align}
The directional antenna gain $G(\theta,\varphi)$ is then given by
\begin{align}
    G(\theta,\varphi) = \lVert \Cm(\theta, \varphi) \rVert^2
\end{align}
which satisfies
\begin{align}\label{eq:atenna_normalization}
    \int_0^{2\pi}\int_0^\pi G(\theta,\varphi)\sin(\theta)d\theta d\varphi = 4\pi \eta_\text{rad}
\end{align}
where $\eta_\text{rad}\in[0,1]$ is the radiation efficiency.
One can define antenna patterns by their directional gain and polarization slant angle $\zeta$, where $\zeta=0$ and $\zeta=\pi/2$ correspond to vertical and horizontal polarization, respectively:
\begin{align}\label{eq:polarization_model}
    \Cm(\theta,\varphi) = \begin{bmatrix} \cos(\zeta) \sqrt{G(\theta, \varphi)} & \sin(\zeta)\sqrt{G(\theta, \varphi)} \end{bmatrix}\tp.
\end{align}
We have chosen this polarization model for its simplicity. Other models are proposed in the literature, e.g., \cite[Sec. 7.3.2]{3gpp38901}.
Closed-form expressions for the directional gain of different antennas can be found in \cite{balanis97} and \cite[Table 7.3-1]{3gpp38901}. In Section~\ref{sec:antenna_pattern}, we describe a parametric differentiable antenna pattern that can be used for gradient-based calibration.

\subsection{Transfer function}
A propagation path consists of a sequence of $Q$ scattering events, such as specular and diffuse reflections, as well as diffraction. For each scattering process, one needs to compute a relationship between the incoming field at the current scattering point $\sv_j\in\RR^3$,  for $j=1,\dots, Q$, and the created far-field at the next scattering point $\sv_{j+1}$, where $\sv_0$ and $\sv_{Q+1}$ denote the position of the transmitter and receiver, respectively.

For every scattering process, the electrical field is represented in its local basis. Since the field is transversal, it can always be represented by two orthogonal components $p$ and $q$. Let us denote by $\Em^{\text{Out}}_{j}(\sv_{{j+1}})\in\CC^2$ the vector of the two components of the outgoing field of the $j$th scattering process at the $j+1$th scattering point, represented in the basis of the $j$th scattering process. The incoming field at the $j+1$th scattering point in its own basis $\Em^{\text{In}}_{j+1}(\sv_{j+1})$ is then given by
\begin{align}\label{eq:rel1}
    \Em^{\text{In}}_{j+1}(\sv_{j+1}) &= \Dm_{j+1} \Em^{\text{Out}}_{j}(\sv_{j+1})
\end{align}
where
\begin{align}\label{eq:basis_transform}
    \Dm_{j+1} = \begin{bmatrix}
                    \LB\hat{\ev}^\text{In}_{p,j+1}\RB\tp\hat{\ev}^\text{Out}_{p,j} & \LB\hat{\ev}^\text{In}_{p,j+1}\RB\tp\hat{\ev}^\text{Out}_{q,j} \\
                    \LB\hat{\ev}^\text{In}_{q,j+1}\RB\tp\hat{\ev}^\text{Out}_{p,j} & \LB\hat{\ev}_{q,j+1}^\text{In}\RB\tp\hat{\ev}^\text{Out}_{q,j}
                \end{bmatrix}
\end{align}
is a basis transformation matrix, and $\LB \hat{\ev}^\text{In}_{p,j}, \hat{\ev}^\text{In}_{q,j}\RB$ and $\LB \hat{\ev}^\text{Out}_{p,j}, \hat{\ev}^\text{Out}_{q,j}\RB$ are pairs of orthogonal basis vectors used for the description of the incoming and outgoing field of the $j$th scattering process, respectively. 

The incoming and outgoing field of the $j$th scattering process are related as
\begin{align}\label{eq:rel2}
    \Em^{\text{Out}}_{j}(\sv_{j+1}) &= \Fm_j \LB \Em^{\text{In}}_{j}(\sv_{j}) \RB
\end{align}
where $\Fm_j: \CC^2\mapsto \CC^2$ depends on the type of scattering process as will be discussed below and, in particular, we have
\begin{align}
    \Em^{\text{In}}_{0}(\sv_{0}) &= \Cm_\text{T}\LB \theta^\text{T}, \varphi^\text{T}\RB\\\label{eq:init}
    \Fm_0 ( \Em^{\text{In}}_{0}(\sv_{0})) &= \frac{e^{-\mathrm{j}2\frac{\pi}{\lambda} d_1}}{d_1} \Em^{\text{In}}_{0}(\sv_{0})
\end{align}
where $d_j = \lVert \sv_j - \sv_{j-1}\rVert$ for $j=1,\dots,Q+1$.

Combining \eqref{eq:rel1}--\eqref{eq:init}, we arrive at the following expression for the relationship between the transmitted and received field
\begin{align}
    \Em^{\text{In}}_{Q+1}(\sv_{Q+1}) &= \Dm_{Q+1}\Fm_Q\LB\cdots\Dm_1\Fm_0\LB\Em^{\text{In}}_{0}(\sv_{0})\RB\RB\\\label{eq:transfer_function}
    &= \Tm \LB \Cm_\text{T}\LB \theta^\text{T}, \varphi^\text{T}\RB\RB e^{-\mathrm{j}2\pi f \tau} 
\end{align}
where we have factored out an exponential term representing the phase shift caused by the total propagation delay ${\tau = (\lambda f)^{-1} \sum_{q} d_q}$. The last equation defines the transfer function $\Tm_i$ in \eqref{eq:cfr_cont} for the considered propagation path.
The structure of this sequence of computations resembles the computational graph of a feed-forward \gls{NN}. As long as all functions $\Fm_j$ are differentiable, so will be the final \gls{CFR}.

\subsection{Reflection}\label{sec:reflection}
We distinguish between specular and diffuse reflections and denote by $S^2$ the fraction of the reflected energy that is diffusely scattered, where $S\in[0,1]$ is the scattering coefficient \cite{degli2007measurement}. Similarly, $R^2$ is the specularly reflected fraction of the reflected energy, where $R=\sqrt{1-S^2}$.
We consider materials that are uniform non-magnetic dielectrics with relative permeability $\mu_r=1$.
A material is described by four parameters, the conductivity $\sigma\ge 0$, the relative permittivity $\varepsilon_r\ge 1$, the scattering coefficient $S$, and the \gls{XPD} coefficient $K_x\in[0,1]$. The latter determines how much energy is transferred into the orthogonal polarization direction after a diffuse reflection \cite{degli2011analysis}. Each material has a scattering pattern $f_s$ that describes, similar to an antenna pattern, the directivity of the scattered field.
The complex relative permittivity $\eta$ is defined as \cite[Eq. (9b)]{itu-r-p2040}
\begin{align}
    \eta = \varepsilon_r - \mathrm{j}\frac{\sigma}{\varepsilon_0 \omega}
\end{align} 
where $\varepsilon_0$ is the vacuum permittivity and $\omega=2\pi f$.

\subsubsection{Specular reflection}
For an incident wave in vacuum with direction $\hat{\kv}_\text{i}$ incident on a surface with permittivity $\eta$ and surface normal $\hat{\nv}$, the field transformation \eqref{eq:rel2} is given as
\begin{align}\label{eq:specular_reflection}
    \Fm_j \LB \Em^{\text{In}}_{j}(\sv_{j}) \RB = \begin{bmatrix}
                                                    R r_\perp & 0\\
                                                    0 & R r_\parallel
                                                 \end{bmatrix} \Em^{\text{In}}_{j}(\sv_{j}) A^\text{r}(\sv_{j+1}, \sv_j) e^{-\mathrm{j}2\frac{\pi}{\lambda}d_{j+1}} 
\end{align}
where $A^\text{r}(\sv_{j+1}, \sv_j)$ is the spreading factor \cite[Eq. (24)]{deschamps1972ray} that depends on the shape of the wavefront and $r_\perp$, $r_\parallel$ are the Fresnel reflection coefficients \cite[Eq. (37)]{itu-r-p2040}:
\begin{align}\label{eq:fresnel-1}
    r_\perp &= \frac{\cos(\theta_\text{i})-\sqrt{\eta -\sin^2(\theta_\text{i})}}{\cos(\theta_\text{i})+\sqrt{\eta -\sin^2(\theta_\text{i})}}\\\label{eq:fresnel-2}
    r_\parallel &= \frac{\eta \cos(\theta_\text{i})-\sqrt{\eta -\sin^2(\theta_\text{i})}}{\eta\cos(\theta_\text{i})+\sqrt{\eta -\sin^2(\theta_\text{i})}}
\end{align}
where $\cos(\theta_\text{i})=-\hat{\kv}_\text{i}\tp\hat{\nv}$. The basis vectors used to compute $\Dm_{j+1}$ can be found in \cite[Eq. (3.3)-(3.8)]{mcnamara1990introduction}. The reflected ray has direction $\hat{\kv}_\text{r}=\hat{\kv}_\text{i}-2(\hat{\kv}_\text{i}\tp\hat{\nv})\hat{\nv}$.
It is obvious that \eqref{eq:specular_reflection} is differentiable \gls{wrt} to the parameters $\varepsilon_r$, $\sigma$, $S$, as well as the input $\Em^{\text{In}}_{j}(\sv_{j})$.

\subsubsection{Diffuse reflection (scattering)}
We consider the model from \cite{degli2011analysis} for which the field transformation \eqref{eq:transfer_function} is given as
\begin{align}\label{eq:scattered_field}
    \Fm_j \LB \Em^{\text{In}}_{j}(\sv_{j}) \RB &= \frac{\lVert \Em^{\text{s}}_{j}(\sv_{j})\rVert}{d_{j+1}} \begin{bmatrix}
        \sqrt{1-K_x} e^{\mathrm{j}\chi_1} \\ \sqrt{K_x}e^{\mathrm{j}\chi_2} 
    \end{bmatrix}e^{-\mathrm{j}2\frac{\pi}{\lambda}d_{j+1}} \\
    \lVert \Em^{\text{s}}_{j}(\sv_{j})\rVert^2 &= \lVert \Em^{\text{In}}_{j}(\sv_{j})\rVert^2 \cos(\theta_\text{i})dA S^2 \Gamma^2 f_\text{s}(\hat{\kv}_\text{i}, \hat{\kv}_\text{s}, \hat{\nv})
\end{align}
where $\chi_1,\chi2\sim U[0,2 \pi)$ are random phase terms, $\hat{\kv}_\text{s}$ is  the direction of the scattered ray, $dA$ is the size of a small area element, $f_\text{s}(\hat{\kv}_\text{i}, \hat{\kv}_\text{s}, \hat{\nv})$ is the scattering pattern, and $\Gamma$ is the fraction of the incoming power that is reflected, i.e.,
\begin{align}
    \Gamma = \left \lVert \begin{bmatrix} r_\perp & 0\\ 0 & r_\parallel \end{bmatrix}  \Em^{\text{In}}_{j}(\sv_{j}) \right\rVert  \lVert  \Em^{\text{In}}_{j}(\sv_{j})\rVert^{-1}.
\end{align}
Under the backscattering lobe model from \cite{degli2007measurement}, the scattering pattern has the following form
\begin{align}
    &f_\text{s}(\hat{\kv}_\text{i}, \hat{\kv}_\text{s}, \hat{\nv}) = \frac1{F_{\alpha_\text{r}, \alpha_\text{s}}(\theta_\text{i})} \nonumber \\\label{eq:backscattering_model}
    & \times \LSB\Lambda\LB\frac{1+\hat{\kv}_\text{r}\tp \hat{\kv}_\text{s}}{2}\RB^{\alpha_\text{r}} + (1-\Lambda)\LB \frac{1+\hat{\kv}_\text{i}\tp \hat{\kv}_\text{s}}{2}\RB^{\alpha_\text{s}}\RSB
\end{align}
where $F_{\alpha_\text{r}, \alpha_\text{s}}(\theta_\text{i})$ normalizes the hemispherical integral of the pattern to one, $\alpha_\text{r},\alpha_\text{s}\ge 1$ are integer parameters that control the lobe width, and $\Lambda\in[0,1]$ determines the fraction of the scattered energy in the specular lobe.
The basis transformation matrix $\Dm_{j+1}$ ensures that energy transfer between both polarization directions is correctly applied.
While \eqref{eq:scattered_field} is differentiable in all material parameters $\varepsilon_r$, $\sigma$, $S$, $K_x$ as well as $\Lambda$, this is not the case for $\alpha_r$ and $\alpha_s$. In Section~\ref{sec:scattering_pattern}, we will present a fully differentiable version of a scattering pattern with similar structure.

\subsection{Diffraction}
Diffraction is modeled according to the \gls{UTD} by \cite{kouyoumjian1974uniform} (see \cite{mcnamara1990introduction} for an introduction)
with the heuristic extension to finitely conducting wedges by \cite{luebbers1984finite} (adopted in \cite{itu-r-p2526}). We consider infinitely long wedges whose two faces can be made of possibly different materials $\eta_0$ and $\eta_1$. Under this model, the field transformation \eqref{eq:rel2} is given as
\begin{align}\label{eq:diffraction}
    \Fm_j \LB \Em^{\text{In}}_{j}(\sv_{j}) \RB &= \Qm\Em^{\text{In}}_{j}(\sv_{j}) A^\text{d}(\sv_{j+1}, \sv_j) e^{-\mathrm{j}2\frac{\pi}{\lambda}d_{j+1}} 
\end{align}
where $\Qm$ is the diffraction matrix and $A^\text{d}(\sv_{j+1}, \sv+j)$ the spreading factor for diffraction. The computation of $\Qm$ is rather involved and requires the computation of Fresnel integrals. However, what matters for the presentation here is that \eqref{eq:diffraction} is differentiable with respect to the parameters of the materials of both faces of the wedge. Diffraction is expressed in an edge-fixed coordinate system \cite[Sec. 6.2]{mcnamara1990introduction}.

\subsection{Path tracing}
We consider a fixed maximum number of specular reflections, possibly followed by a diffuse reflection. Only first-order diffraction (i.e., transmitter-diffraction-receiver) is assumed and refraction ignored.
Specular paths are found by first determining a set of candidate primitives using shooting-and-bouncing rays (SBR) which are then corrected to exact specular chains using the image method. Diffuse reflections are found through next event estimation in the same process. Candidate edges for diffraction are also identified during SBR and valid points of diffraction are found by the principle of the shortest distance \cite{keller1962geometrical}.
Note that we do not seek to compute gradients \gls{wrt} the scene geometry, although this would be in principle possible, e.g., \cite{inv_rendering}. This has the advantage that the calibration is carried out on the field computations alone which avoids re-tracing of paths for every training iteration.

%% file: sections/diff-rt.tex
\section{Differentiable ray tracing for scene calibration}\label{sec:diff_rt}
In this section, we will explain how one can parametrize antenna and scattering patterns as well as material properties in a suitable way for gradient-based optimization.

\subsection{Trainable antenna patterns}\label{sec:antenna_pattern}
Antenna patterns as described in Section~\ref{sec:antenna_modeling} are square integrable functions on the sphere that need to satisfy the normalization constraint \eqref{eq:atenna_normalization}. We will now propose a parametric model for a trainable antenna pattern that has these properties. We restrict ourselves to trainable directional gains $G(\theta,\varphi)$ from which the antenna pattern can be computed according to \eqref{eq:polarization_model}. Since many antenna patterns have strong directional characteristics, we model them as a mixture of $M$ spherical Gaussians (or von Mises-Fisher distributions):
\begin{align}
    G(\theta, \varphi) &= \eta_\text{rad}\sum_{i=1}^M \frac{w_i}{a_i} e^{\lambda_i \muv_i\tp \hat{\rv}(\theta, \varphi)}\\\label{eq:normalization_sphere}
    a_i &= \frac{2\pi}{\lambda_i}\LB e^{\lambda_i} - e^{-\lambda_i}\RB
\end{align}
where $\eta_\text{rad}\in[0,1]$ is the radiation efficiency, $w_i\in[0,1]$ are mixture weights satisfying ${\sum_i w_i=1}$, $\lambda_i> 0$ are concentration parameters, and $\muv_i\in\RR^3$ are unit norm mean directions. All of these quantities are modeled as trainable variables.

\subsection{Trainable scattering patterns}\label{sec:scattering_pattern}
In contrast to antenna patterns, scattering patterns are defined on the hemisphere and take three arguments, namely the direction of the incoming and outgoing rays, as well as the surface normal. Both differences make their parametrization more difficult. We propose the following model which can be considered the hemispherical Gaussian-equivalent to \eqref{eq:backscattering_model} with the addition of a diffuse component:
\begin{align}\label{eq:trainable_scattering_pattern}
    f_\text{s}\LB \hat{\kv}_\text{i}, \hat{\kv}_\text{s}, \hat{\nv} \RB = \frac{w_1}{\pi}\hat{\kv}_\text{s}\tp\hat{\nv} + \frac{w_2}{a_2} e^{\lambda_2 \hat{\kv}_\text{s}\tp\hat{\kv}_\text{i}} + \frac{w_3}{a_3} e^{\lambda_3 \hat{\kv}_\text{s}\tp\hat{\kv}_\text{r}}
\end{align}
where $\lambda_2, \lambda_3\ge0$ are concentration parameters, $w_i\in[0,1]$ are mixture weights satisfying ${\sum_i w_i=1}$, $\hat{\kv}_\text{r}= \hat{\kv}_\text{i}-2(\hat{\nv}\tp\hat{\kv}_\text{i})\hat{\nv} $ is the direction of the specular reflection, and $a_i$, $i=2,3$, are approximate normalization factors given by \cite[Algorithm 1]{meder2018hemispherical}
\begin{align}\label{eq:normalization_hemisphere}
    &a_i = \frac{2\pi}{\lambda_i}\LB e^{-\lambda_i}-1\RB\LB 2s_i-1\RB \\
    &s_i = \frac{e^{t_i} e^{t_i\cos(\beta_i)} - 1}{\LB e^{t_i}-1\RB\LB e^{t_i \cos(\beta_i)}+1\RB}\\
    &t_i = \sqrt{\lambda_i}\frac{1.6988\lambda_i^2 + 10.8438\lambda_i}{\lambda_i^2+6.2201\lambda_i + 10.2415}\\
    &\cos(\beta_2) = \hat{\kv}_\text{s}\tp\hat{\kv}_\text{i}\\
    &\cos(\beta_3) = \hat{\kv}_\text{s}\tp\hat{\kv}_\text{r}.
\end{align}
The difference between the normalization factors $a_i$ in \eqref{eq:normalization_sphere} and \eqref{eq:normalization_hemisphere} arises because the integral of the hemispherical Gaussian has no closed-form solution for arbitrary values of $\cos(\beta_i)$. The authors of \cite{meder2018hemispherical} propose a heuristic approximation which we have used here that achieved excellent accuracy in our experiments, is differentiable, and easy to implement.

\subsection{Trainable materials}\label{sec:materials}
As described in Section~\ref{sec:reflection}, a material is characterized by its constituent parameters $\sigma\ge0$ and $\varepsilon_r\ge 1$, the scattering coefficient $S\in[0,1]$, and the \gls{XPD} coefficient $K_x\in[0,1]$. Although one could immediately consider these parameters as trainable variables, it turned out to be beneficial to use an over-parametrization that helps avoid local minima during gradient-based optimization \cite{pmlr-v38-choromanska15}. We adopt the following model:
\begin{align}\label{eq:trainable_materials_1}
    \sigma &= e^{\vv\tp\wv_1} \\
    \varepsilon_r &= 1 + e^{\vv\tp\wv_2} \\
    S &= \sigmoid(\vv\tp\wv_3) \\\label{eq:trainable_materials_4}
    K_x &= \sigmoid(\vv\tp\wv_4)
\end{align}
where $\sigmoid(x)=1/(1+e^{-x})$, $\vv\in\RR^L$ and $\wv_i\in\RR^L$ for $i=1,\dots,4$, are trainable vectors in $L$ dimensions. The \emph{read-out vector} $\vv$ is used to convert the \emph{embeddings} $\wv_i$ to scalar values. As $\sigma$ and $\varepsilon_r$ need to cover a very large range of values, they are represented in the logarithmic domain.

\subsection{Neural materials}\label{sec:neural_materials}
The granularity at which material properties in a scene are modeled is an important part of ray tracer calibration. In the simplest case, all objects are made of the same material. One can then subdivide the scene into different object classes with individual materials, e.g., floor, walls, and ceiling. In the most extreme case, all triangles of the meshes defining objects in the scene would have their own materials. This discretization has two difficulties. First, the number of materials that need to be calibrated can become very large. Second, the process of discretizing a scene into objects or parts of objects sharing the same material cannot be easily automated. For this reason, we propose a different approach here called \emph{neural materials} that does not require discretization, is easy to implement and calibrate, and can, in principle, scale to arbitrarily large scenes.

\begin{figure}
    \centering
    \includegraphics[width=\columnwidth]{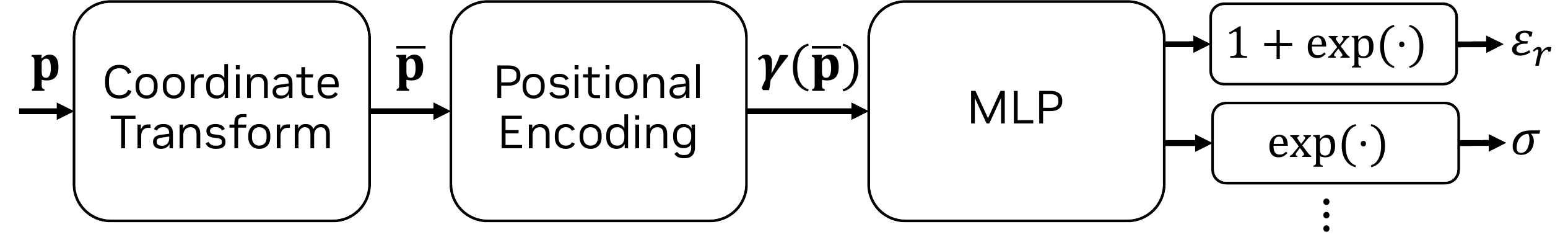}
    \caption{Processing steps of the computation of neural materials}
    \label{fig:neural_materials}
\end{figure}

The key idea is to have a small \gls{NN}, such as a \gls{MLP}, that computes the  corresponding material properties for every ray-scene intersection point. This does not need to be limited to $\varepsilon_r$, $\sigma$, $S$, and $K_x$, but may also comprise the computation of the weights and concentration parameters of the scattering pattern in \eqref{eq:trainable_scattering_pattern} among others. 
Rather than having the \gls{NN} operate directly on the input coordinates $\pv=[x,y,z]\tp\in\RR^3$, we adopt the positional encoding \cite{mildenhall2021nerf}
\begin{align}\label{eq:pos_encoding}
    \hspace{-5pt}\gamma(p) = \LSB \sin(2\pi p), \cos(2\pi p),\dots,\sin(2^{L}\pi p), \cos(2^{L}\pi p)\RSB
\end{align}
that is separately applied to each coordinate. This has been essential to enable \glspl{NN} to better represent functions with high frequency components over low-dimensional domains \cite{NEURIPS2020_55053683}. The parameter $L$ controls how fast material properties can vary across space. To make the positional encoding unique, the input coordinates are transformed to the unit cube:
\begin{align}
    \bar{\pv} = \frac{\pv-\bv}{\Delta_b}
\end{align}
where $\Delta_b\in\RR$ is the length and $\bv\in\RR^3$ the center of the \gls{AABB} of the scene, which 
defines the smallest bounding box that encloses all objects in the scene and has edges parallel to the coordinate axes. The normalized and encoded coordinates $\gamma(\bar{\pv})\in\RR^{6L}$ are fed into an \gls{MLP} which has one output for each parameter. The outputs are then transformed as in \eqref{eq:trainable_materials_1}--\eqref{eq:trainable_materials_4}. Fig.~\ref{fig:neural_materials} shows a block diagram of all processing steps. One may also apply the positional encoding on the coordinates represented in the UV space \cite{pbrbook} or replace the positional encoding by a multiresolution hash encoding as in \cite{mueller2022instant}.

\subsection{Loss function}
One challenge related to the calibration of a ray tracer to measurements is the choice of metric or loss function. In our experiments, we have used the following per-example loss
\begin{align}\label{eq:loss}
    \mathcal{L} = l( \tau_\text{RMS}, \hat{\tau}_\text{RMS}) + l( P, \hat{P})
\end{align}
where 
\begin{align}\label{eq:smape}
    l\LB x, y\RB = \frac{\lvert x-y\rvert}{x+ y} \in[0,1]
\end{align}
is the \gls{SMAPE},
\begin{align}\label{eq:delay_spread}
    \tau_\text{RMS} &= \sqrt{\sum_{\ell} \LB \frac{\ell-\bar{\tau}}{W}\RB^2 \frac{p_\ell}{P}}
\end{align}
is the \gls{RMS} delay spread with $\bar{\tau} = \sum_{\ell} \ell \frac{p_\ell}{P}$ and $p_\ell= \left|h[\ell]\right|^2$, and $P$ is the total channel gain: 
\begin{align}\label{eq:channel_gain}
    P = \sum_{\ell} p_\ell.
\end{align}

The values $\hat{P}$ and $\hat{\tau}_\text{RMS}$ are obtained from ray-traced \glspl{CIR}, while $P$ and $\tau_\text{RMS}$ are obtained from measurements. The intuition behind the loss in \eqref{eq:loss} is that the first term measures differences in the delay distribution of the relative path powers while the second term measures differences in the absolute power. As the loss is typically computed over a batch of measurements taken at different positions in a scene, the normalization in \eqref{eq:smape} ensures that all examples are given equal importance. Without this normalization, the calibration procedure would focus on areas with strong channels, leading to poor performance in areas of weak coverage. The reason why we have chosen this loss over alternatives, such as the power-angular-delay spectrum \cite{charbonnier2020calibration}, is that both the \gls{RMS} delay spread and absolute power are agnostic to time offsets which are present in our dataset. However, alternative losses may be used as drop-in replacements for calibration with other measurements where this is not an issue.

\subsection{Scaling of measurements}\label{sec:scaling}
A particularity of our channel measurements is that no absolute power reference is known. This means that all examples of the measured dataset are scaled by a common but unknown value $\alpha\in\RR$. In order to find a suitable scaling parameter, we estimate it iteratively by an \gls{EMA} during the calibration process:
\begin{align}\label{eq:scaling_param}
    \alpha_i = \delta \alpha_{i-1} + (1-\delta)\hat{\alpha}_i,\quad i=1,2,\dots
\end{align}
where $B$ is the batch size, $\delta>0$ the decay parameter, ${\alpha_0=\hat{\alpha}_1}$, and $\hat{\alpha}_i$ is computed as
\begin{align}\label{eq:alpha_hat}
    \hat{\alpha}_i = \arg\min_\alpha \frac{1}{B}\sum_{b=1}^B \LB \alpha P_b - \hat{P}_b \RB^2 = \frac{\sum_b P_b \hat{P}_b}{\sum_b P_b^2}.
\end{align}
In the limit, $\alpha_i$ is expected to be close to the scaling parameter that solves \eqref{eq:alpha_hat} for the entire dataset. Computing the latter at every iteration would be too costly. In our experiments, this approach has led to better results than learning a scaling factor via gradient descent.

\subsection{Training algorithm}
Similar to training an \gls{NN}, we calibrate our scene using \gls{SGD} on the loss \eqref{eq:loss} \gls{wrt} to all trainable parameters of the materials, antenna and scattering patterns described in the previous sections and which we denote by $\Omegam$. It is assumed that a reference dataset of $D$ \glspl{CIR} $\hv_d$ together with their associated positions $\pv_d$ is available. The algorithm starts by randomly initializing all trainable parameters $\Omegam$. It then iteratively computes the scaling factor $\alpha_i$ and the ray tracing predictions $\hat{\hv}_b =\mathop{\textsf{RT}}(\pv_b;\Omegam)$ for a random batch of $B$ examples and uses the gradient of the average loss over the batch to update the parameters until convergence. Algorithm~\ref{alg:training} provides a formal description of this process using vanilla \gls{SGD} with a learning rate $\beta$ for the sake of compactness. Note that any other \gls{SGD}-variant might be used as well: We employ Adam \cite{adam} throughout Section~\ref{sec:experiments}.

\begin{algorithm}
    \caption{Gradient-based Scene Calibration}\label{alg:training}
    \KwIn{$B$, $\delta$, $\beta$}
    \KwData{$(\hv_d, \pv_d)$ for $d\in\Dc=\{1,\dots,D\}$}
    Randomly initialize all trainable parameters $\Omegam$\\
    $i \gets 1$\\
    \Repeat{convergence}{
      Draw random batch of $B$ examples $\Bc\subseteq\Dc$\\
      Compute $\alpha_i$ as in \eqref{eq:scaling_param}\\
      \For{$b\in\Bc$}{
          $\hv_b \gets \sqrt{\alpha_i}\hv_b$\\
          $\hat{\hv}_b \gets \mathop{\textsf{RT}}(\pv_b;\Omegam)$\\
          Compute sample loss $\Lc_b$ as in \eqref{eq:loss}
      }
      $\Lc = \frac{1}{B}\sum_b \Lc_b$\\
      $\Omegam \gets \Omegam - \beta \nabla_\Omegam \Lc $\\
      $i \gets i+1$
    }
\end{algorithm}

%% file: sections/experiments.tex
\section{Experiments}\label{sec:experiments}

\subsection{Scenario and measurements}

\begin{figure*}[th]
  \centering
  \includegraphics[width=\textwidth]{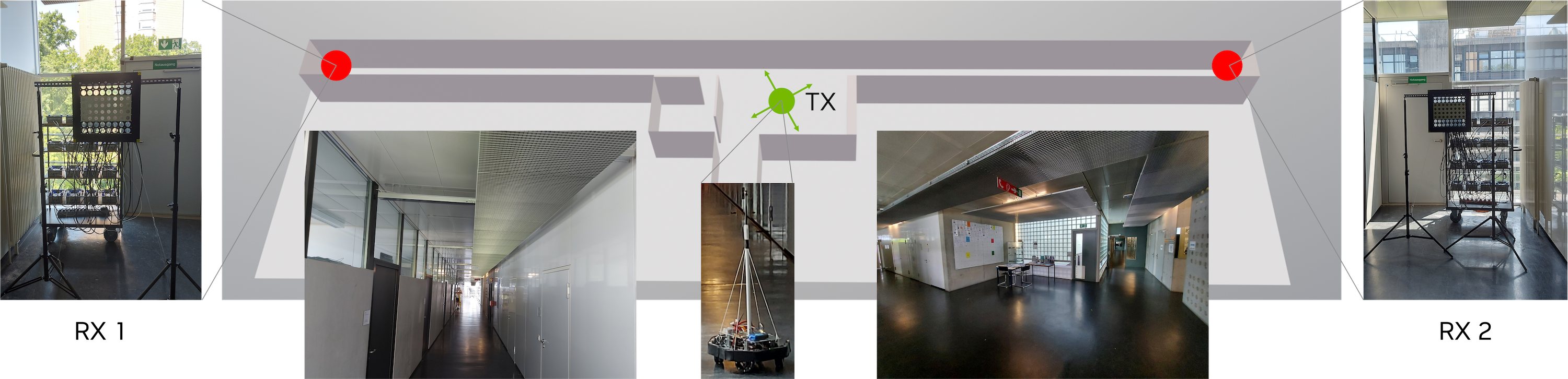}
  \caption{Measurement setup within the University of Stuttgart's Institute of Telecommunications. Two 32-antenna panels from the DICHASUS channel sounder were positioned at opposing ends of the main corridor. A remotely-operated transmitter navigated the hallway and entrance hall along a pseudo-random path.
  \label{fig:scenario}}
\end{figure*}

The calibration dataset stems from a small subset of a measurement campaign \cite[``dichasus-dc01'']{dichasus_dataset} carried out at the Institute of Telecommunications of the  University of Stuttgart, utilizing the DICHASUS channel sounder \cite{dichasus2021}. Instructions about how to access the dataset and 3D scene model can be found in \cite{diff-rt-calibration}. Measurements were executed at a center frequency of ${f=\SI{3.438}{\giga\hertz}}$ and a signal bandwidth of \SI{50}{\mega\hertz}. \Glspl{CFR} across \num{1024} \gls{OFDM} subcarriers were captured at around \num{32.5}k measurement positions from a remote-controlled single-antenna dipole transmitter
to two distributed uniform planar \num{8}x\num{4} antenna panels with $\lambda/2$ spacing and  vertically polarized antennas. The panels were located at opposing ends of the institute's hallway (around \SI{50}{\meter} apart) and the transmitter had its location meticulously tracked  using a tachymeter. The data collection was done within the hallway and the entrance hall, as illustrated in Fig.~\ref{fig:scenario}. Despite the intricate nature of this setting, characterized by composite walls of drywall, glass, and metal, a ceiling half-covered by a metal grid, and a sizable magnetic board in the entrance hall, we opted for a rudimentary handcrafted 3D model. This model solely represents the visible walls, floor, and ceiling.

\subsection{Synthetic dataset}
To validate that scene parameters can be accurately recovered using differentiable ray tracing combined with gradient descent-based optimization, we produced a synthetic dataset using the 3D model of our environment. To make things simple, we considered only a single receive antenna located in the center of the RX~2 panel (see Fig.~\ref{fig:scenario}). We then computed \glspl{CFR} through ray tracing for 256 random transmitter positions, taking into account up to three specular reflections as well as first-order diffuse scattering and diffraction. All materials were assumed to adhere to the backscattering lobe model with $\alpha_r=5$, $\alpha_i=8$, and $\Lambda=0.8$. Other material parameters are detailed in Table~\ref{tab:synthetic_params}. In contrast to \eqref{eq:scattered_field}, we did not apply random phase shifts to diffusely reflected paths, i.e., $\chi_1=\chi_2=0$.

We then pretended that the material parameters as well as the scattering and transmit antenna patterns were unknown and tried to recover them by gradient descent on the loss function in \eqref{eq:loss} using the trainable parametrizations described in Sections~\ref{sec:antenna_pattern}--\ref{sec:materials}. The material embeddings were chosen to have $L=30$ dimensions and the transmit antenna pattern $M=3$ mixtures. It was not necessary to learn a scaling parameter which was set to $\alpha=1$ (see Section~\ref{sec:scaling}). To speed-up the training process, we separated the path tracing from the field computation. In each training iteration, we computed the field and resulting \glspl{CFR} for the precomputed geometrical paths and updated all trainable weights based on the loss using the Adam optimizer with a learning rate of $0.01$.

\begin{figure*}[t]
    \centering
    \begin{subfigure}{0.49\textwidth}
      \includegraphics[width=\linewidth]{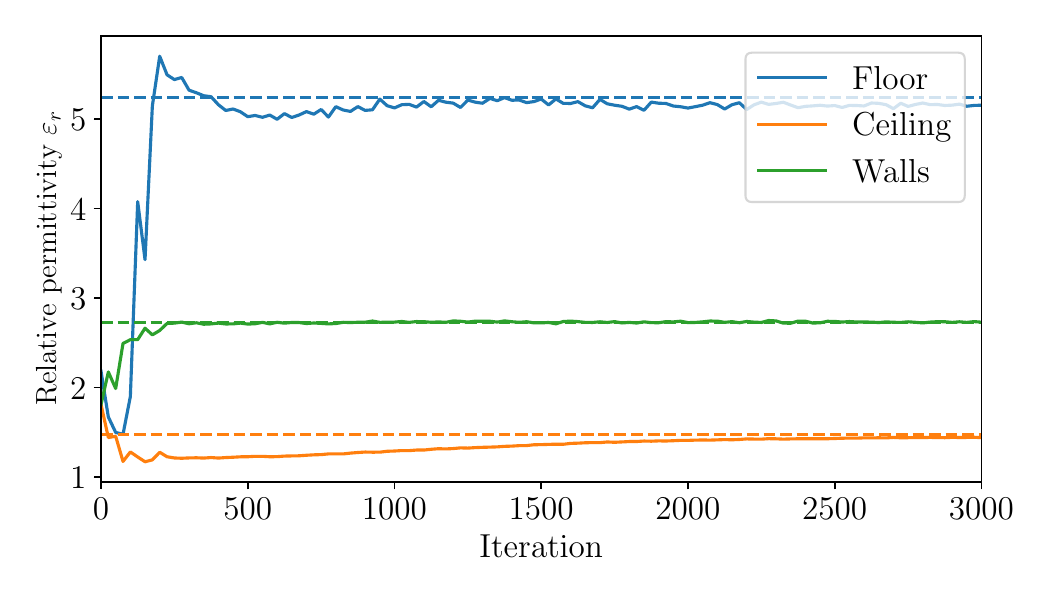}
      \caption{}
    \end{subfigure}
    \hfill
    \begin{subfigure}{0.49\textwidth}
      \includegraphics[width=\linewidth]{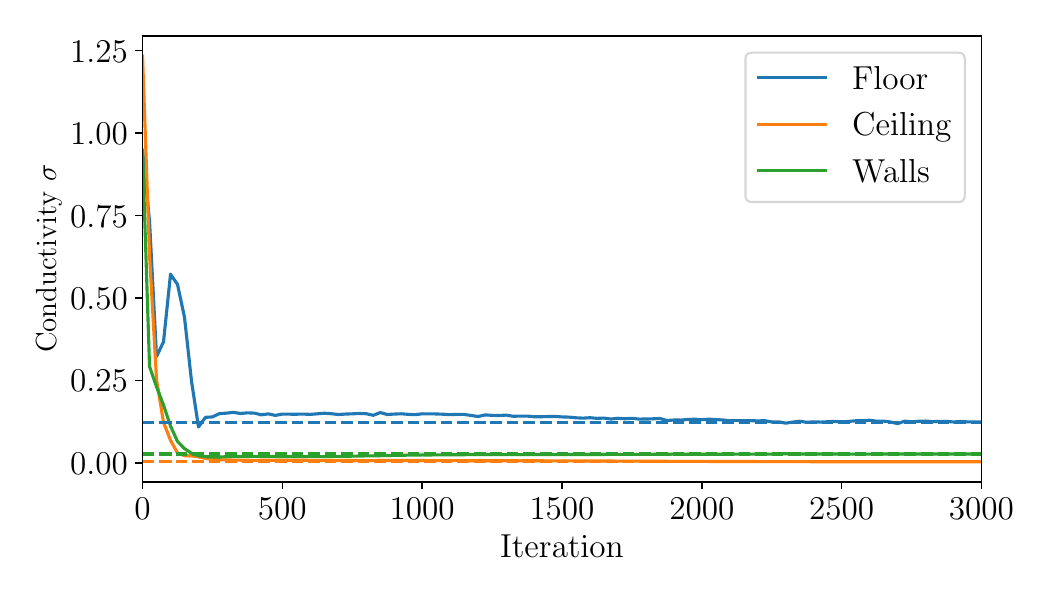}
      \caption{}
    \end{subfigure}
    
    \begin{subfigure}{0.49\textwidth}
      \includegraphics[width=\linewidth]{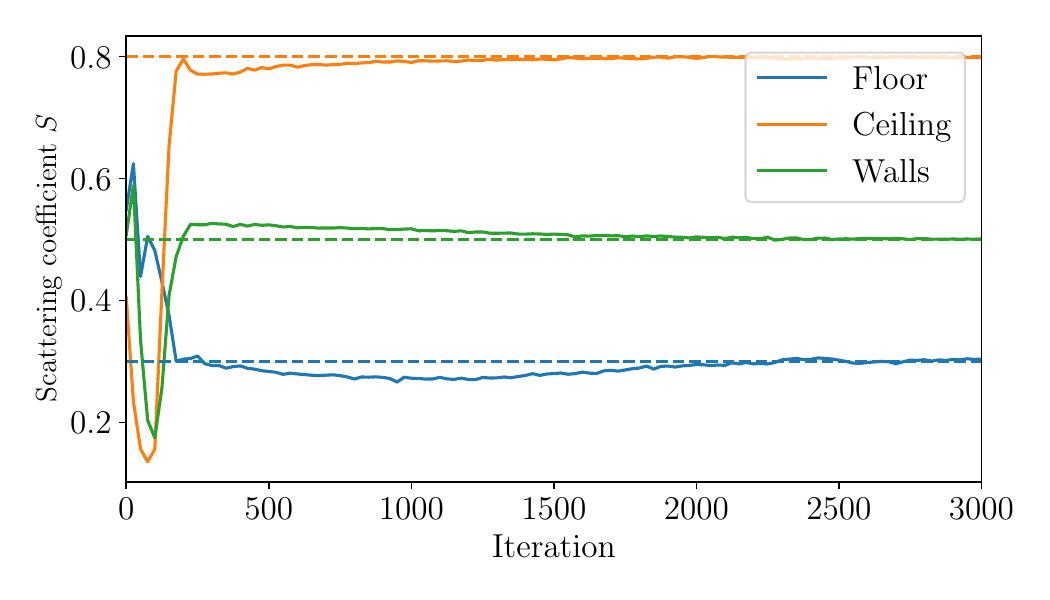}
      \caption{}
    \end{subfigure}
    \hfill
    \begin{subfigure}{0.49\textwidth}
      \includegraphics[width=\linewidth]{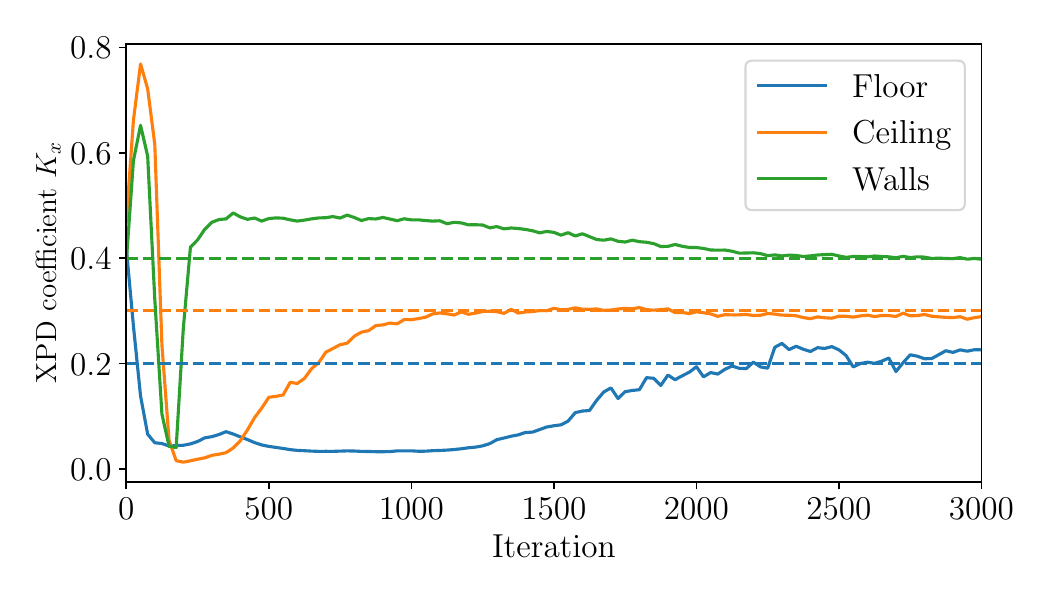}
      \caption{}
    \end{subfigure}
    \caption{Learning curves for the material parameters of all objects in the scene. Dashed and solid lines represent ground truth and learned values, respectively. All parameters can be almost perfectly recovered within \num{3000} iterations.}
    \label{fig:learned_materials}
\end{figure*}

\begin{figure}
    \centering
    \begin{subfigure}{0.49\columnwidth}
      \centering
        \includegraphics[width=\columnwidth]{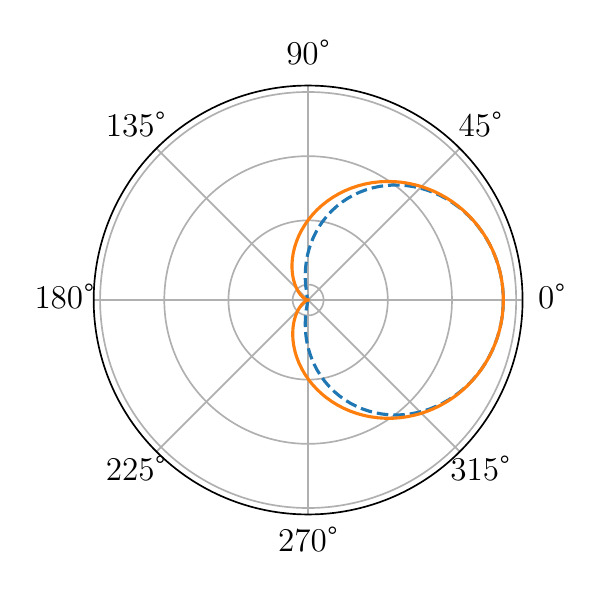}
        \caption{Horizontal cut}
      \end{subfigure}
      \hfill
    \begin{subfigure}{0.49\columnwidth}
        \centering
        \includegraphics[width=\columnwidth]{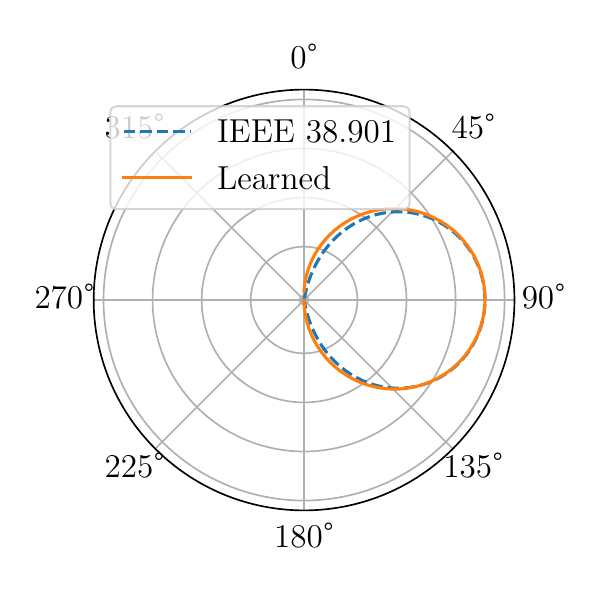}
        \caption{Vertical cut}
      \end{subfigure}
      \caption{Visualization of the learned antenna pattern}
      \label{fig:learned_antenna_pattern}
\end{figure}

\begin{figure}
    \centering
    \includegraphics[width=0.8\columnwidth, trim=0 70pt 0 0]{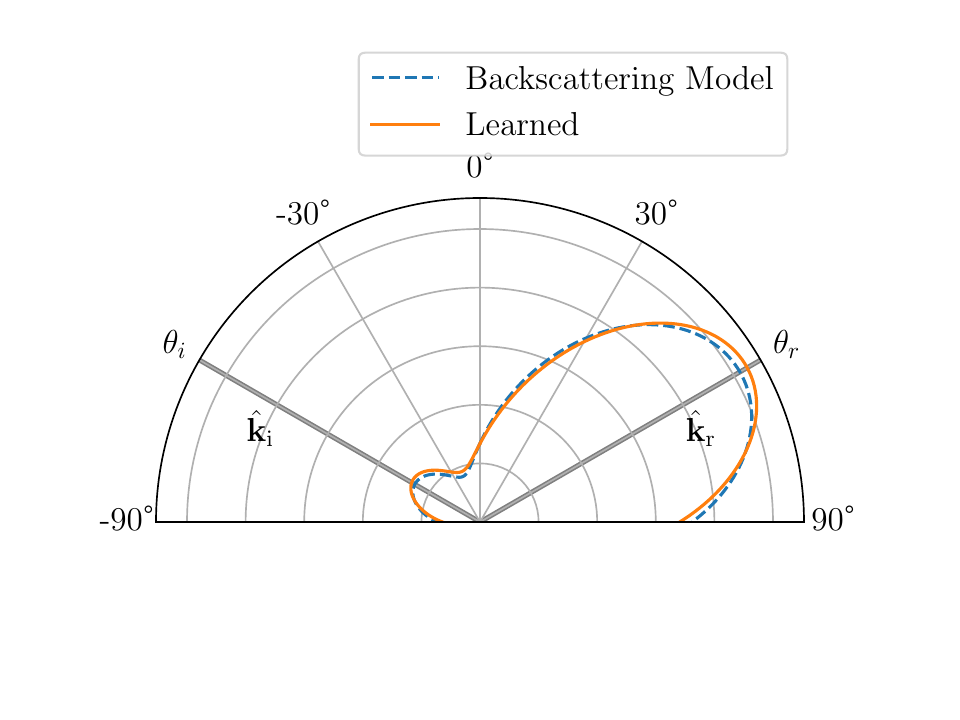}
    \caption{Visualization of the learned backscattering pattern.
    The quantities $\theta_\text{i}$ and $\theta_\text{r}$ denote the angles of the incident and specular directions with the surface normal, respectively.}
    \label{fig:learned_scattering_pattern}
\end{figure}

Fig.~\ref{fig:learned_materials} shows learning curves for the four different material parameters of all three objects in the scene. After around \num{3000} training steps (approx.\ \SI{10}{\minute} on an NVDIA RTX 3090), all parameters have converged to the correct values. Fig.~\ref{fig:learned_antenna_pattern} and Fig.~\ref{fig:learned_scattering_pattern} visualize the learned antenna and scattering patterns. Both match the ground truth also very closely. Overall, these results demonstrate that it is possible to recover all scene parameters via gradient descent from an ideal synthetic dataset.

\begin{table}
    \caption{Scene parameters for synthetic dataset generation}
    \label{tab:synthetic_params}
    \centering
    \begin{tabular}{|l|c|c|c|c|}
    \hline
    Object                      & $\varepsilon_r$ & $\sigma$ & $S$ & $K_x$\\
    \hline
    Floor (ITU Concrete \cite{itu-r-p2040}) & 5.24 & 0.121 & 0.3 & 0.2 \\
    \hline
    Walls (ITU Plasterboard)    & 2.73 & 0.027 & 0.5 & 0.4 \\
    \hline
    Ceiling (ITU Ceiling board) & 1.48 & 0.004 & 0.8 & 0.3 \\
    \hline
    \end{tabular}
\end{table}

\subsection{Calibration with measured data}\label{sec:measurements}
\begin{figure*}[t]
  \centering
  \includegraphics[width=\textwidth]{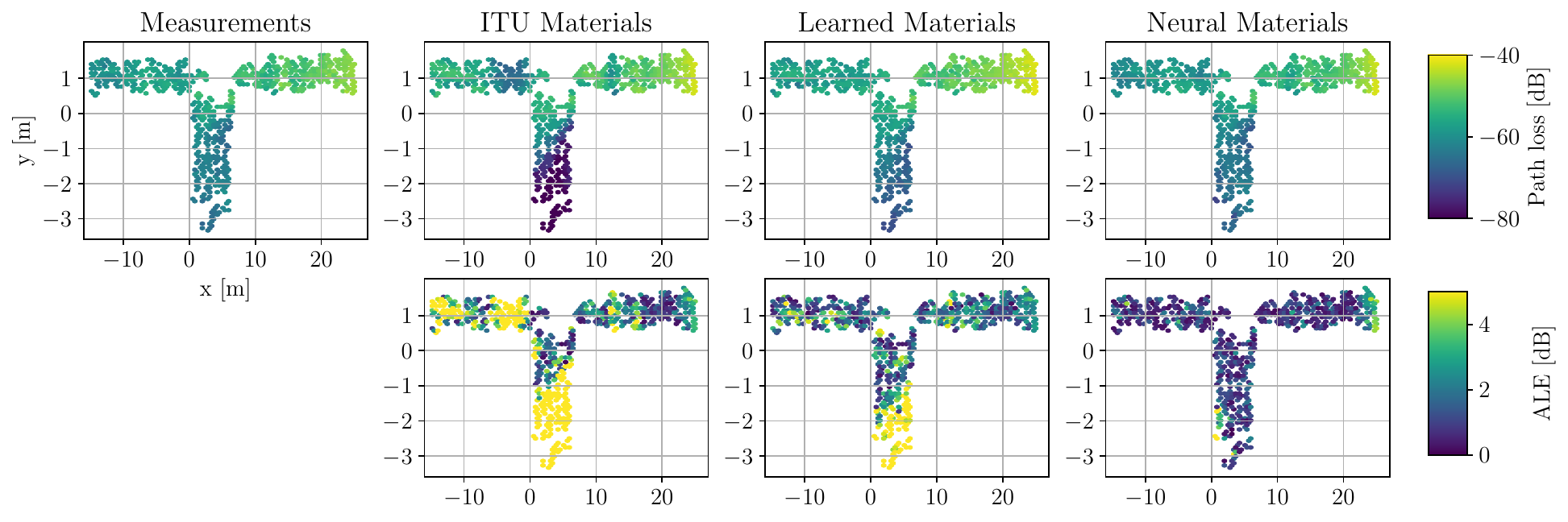}
  \caption{Visualization of path loss ($-P_\text{dB}$) and absolute logarithmic error (ALE) across space for different calibration techniques}
  \label{fig:path_loss}
\end{figure*}

Calibrating our ray tracer to the measured dataset faces several challenges. As mentioned in Section~\ref{sec:scaling}, all measured \glspl{CFR} are scaled by a common but unknown factor $\alpha$ which needs to be estimated. This scaling factor has a very significant impact on the material properties. For example, if the former is too large, the radio environment needs to be overly ``reflective'' to compensate for the underestimated path loss. Second, no information about absolute path delays is available. For this reason, all measured \glspl{CIR} are aligned to have the first significant peak on tap $\ell=0$. 

Although we may in principle calibrate antenna and scattering patterns as done for the synthetic dataset, we did not observe any benefits in doing so for our measured dataset. For this reason, diffuse reflections were turned off and the receive and transmit antennas were respectively configured with vertically polarized 3GPP 38.901 \cite{3gpp38901} and ideal dipole antenna patterns. We have simulated specular reflections up to fifth order as well as first order diffractions.
To keep the size of the dataset manageable, we have only used \num{10}k of the available \num{32.5}k measurement points. 
These were split randomly into \num{5000} for training, \num{100} for validation, and \num{4900} for testing.

The calibration accuracy is quantified as follows. At every measurement point, we compute the average total channel gain $P$ \eqref{eq:channel_gain} and \gls{RMS} delay spread $\tau_\text{RMS}$ \eqref{eq:delay_spread} over the \glspl{CIR} from the transmitter to all \num{64} receive antennas across the two arrays. These quantities are denoted by $\bar{P},\bar{\tau}_\text{RMS}$ and $\hat{\bar{P}},\hat{\bar{\tau}}_\text{RMS}$ for the measured and ray-traced \glspl{CIR}, respectively. We then compute the absolute logarithmic error (ALE) for the channel gain:
\begin{align}
  \text{ALE} = \lvert \bar{P}_\text{dB} -  \hat{\bar{P}}_\text{dB} \rvert
\end{align}
and the relative absolute error (RAE) for the delay spread:
\begin{align}
  \text{RAE} = \frac{\lvert \bar{\tau}_\text{RMS} -  \hat{\bar{\tau}}_\text{RMS}\rvert}{\bar{\tau}_\text{RMS}}.
\end{align}

In the following, we will show results for an uncalibrated baseline as well as two different calibration techniques.
For the baseline that will be referred to as ``ITU Materials'', the scene is configured as described in Table~\ref{tab:synthetic_params} with $S=K_x=0$. The first calibration technique is named ``Learned Materials'' and refers to the parametrization described in Section~\ref{sec:materials} with material embeddings of $L=30$ dimensions. The second, called ``Neural Materials'', is based on the approach described in Section~\ref{sec:neural_materials} with a positional encoding of size $L=10$ and an \gls{MLP} consisting of four hidden layers with \num{128} neurons each and \gls{ReLU} activations.

\begin{figure}
  \centering
  \begin{subfigure}{\columnwidth}
      \includegraphics[width=\columnwidth]{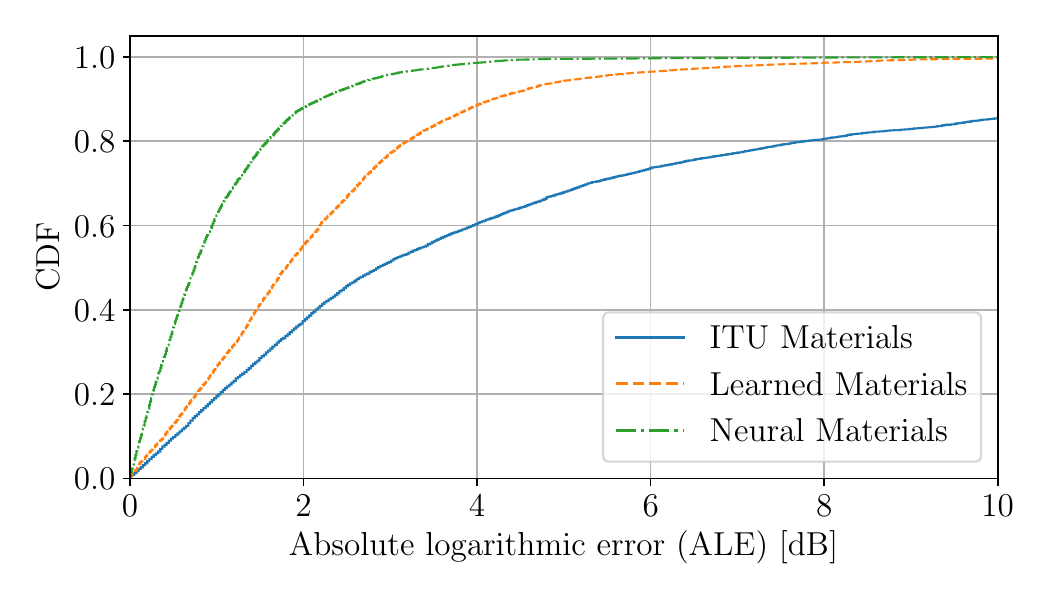}
      \caption{Channel gain}
      \label{fig:cdf_channel_gain}
    \end{subfigure}
    \hfill
    \begin{subfigure}{\columnwidth}
      \includegraphics[width=\columnwidth]{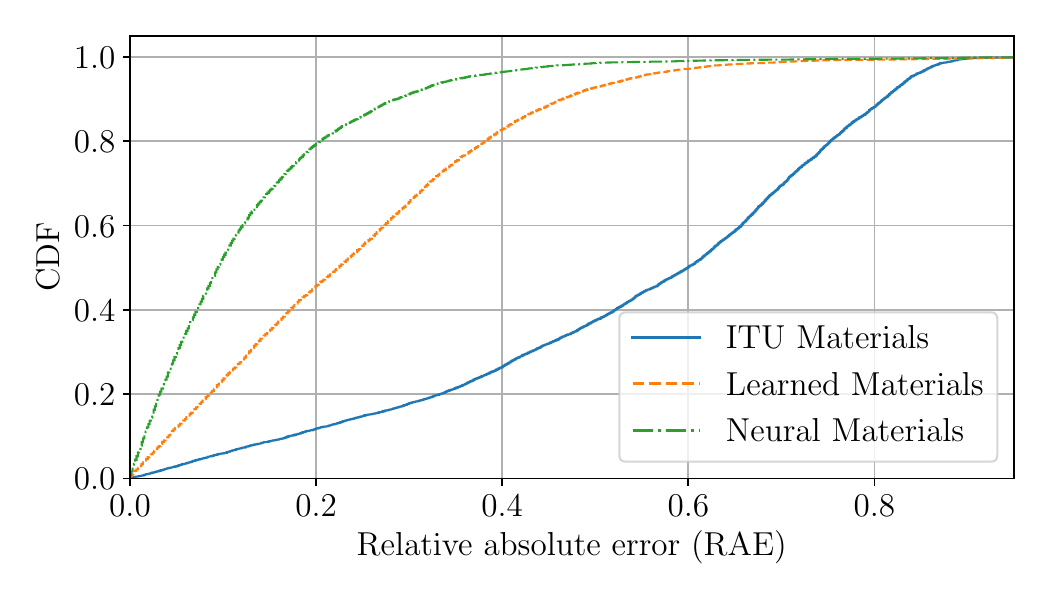}
      \caption{Delay spread}
      \label{fig:cdf_delay_spread}
    \end{subfigure}
    \caption{\Glspl{CDF} of the ALE and RAE}
    \label{fig:error_cdfs}
\end{figure}

The top row of Fig.~\ref{fig:path_loss} shows a heatmap of the path loss $-P_\text{dB}$, i.e., the inverse total channel gain, for the measurements, the ``ITU Materials'', as well as both calibration techniques for RX~2, located in the right wing. The bottom row shows the associated ALE. One can clearly see the benefits of ``Learned Materials'' in the left corridor where the ALE is substantially improved \gls{wrt} to the ``ITU Materials''. However, due to the overly simplistic scene modelling, large errors in the entrance hall persist even after calibration. The ``Neural Materials'' show uniformly improved performance across the whole area. Fig.~\ref{fig:cdf_channel_gain} provides another look at these results via the \gls{CDF} of the ALE. ``Neural Materials'' not only achieve superior accuracy for the path loss, but also the delay spread
which can be seen from the \gls{CDF} of the RAE in Fig.~\ref{fig:cdf_delay_spread}. The corresponding error statistics are provided in Table~\ref{tab:error_stat}

\begin{table}
  \centering
  \caption{Error statistics}
  \label{tab:error_stat}
  \begin{tabular}{|c||c|c|c|c|}
  \hline
  \multicolumn{1}{|c||}{Metric} & \multicolumn{2}{|c|}{ALE (power)} & \multicolumn{2}{c|}{RAE (delay spread)} \\ \hline\hline
     Statistic       & mean       & std         & mean       & std \\ \hline \hline
   ITU Materials     & \num{4.93} & \num{5.26} & \num{0.54} & \num{0.23} \\ \hline
   Learned Materials & \num{2.16} & \num{1.69}  & \num{0.24} & \num{0.17} \\ \hline
   Neural Materials  & \num{1.00} & \num{0.95} & \num{0.13} & \num{0.12} \\ \hline
  \end{tabular}
\end{table}
 
\begin{figure*}[t]
  \centering
  \begin{subfigure}{0.48\textwidth}
    \includegraphics[width=\linewidth]{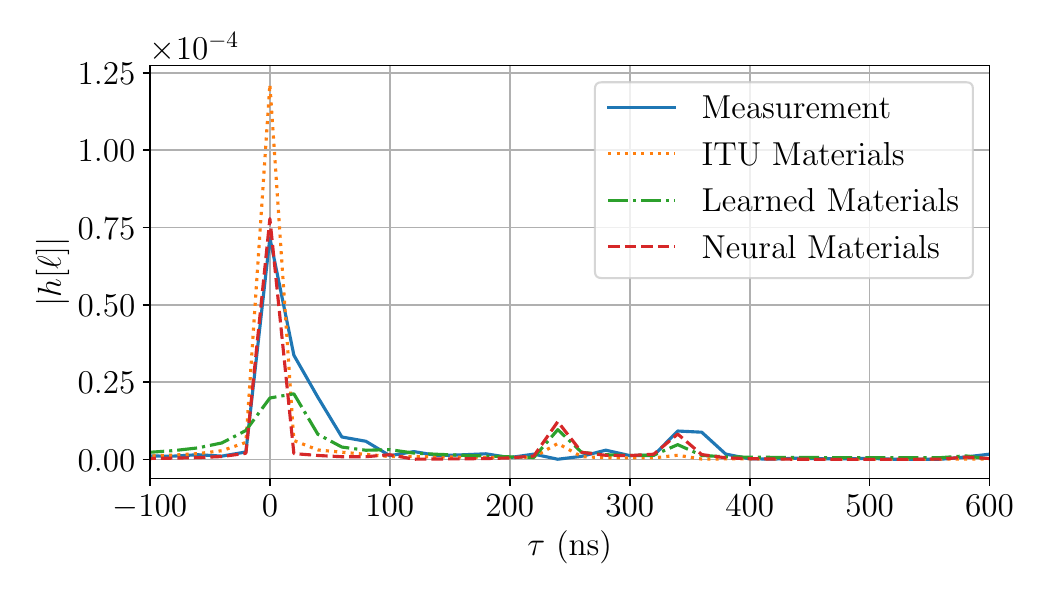}
  \end{subfigure}
  \hfill
  \begin{subfigure}{0.48\textwidth}
    \includegraphics[width=\linewidth]{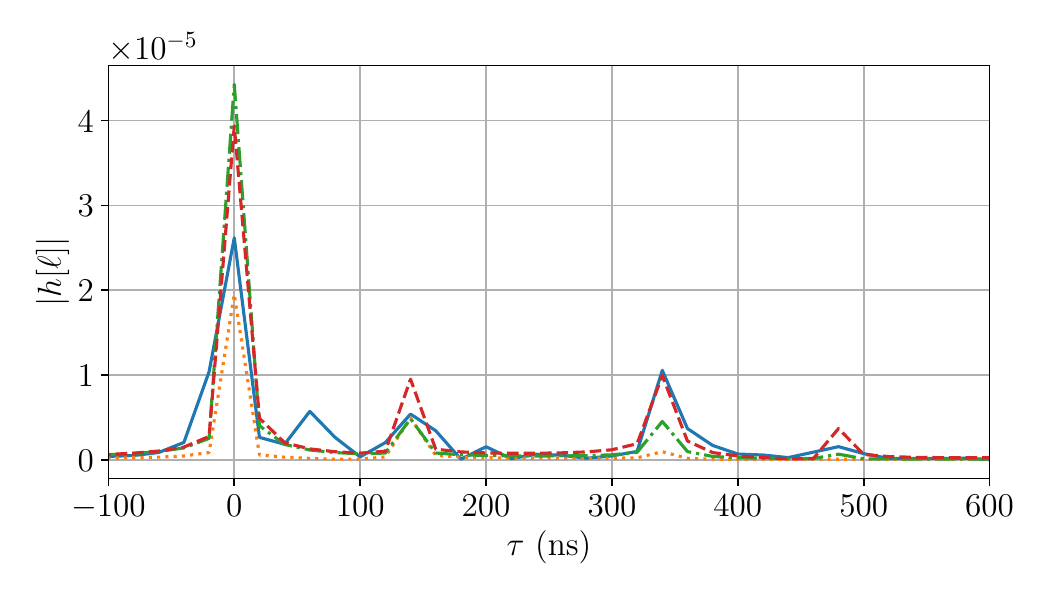}
  \end{subfigure}
  \begin{subfigure}{0.48\textwidth}
    \includegraphics[width=\linewidth]{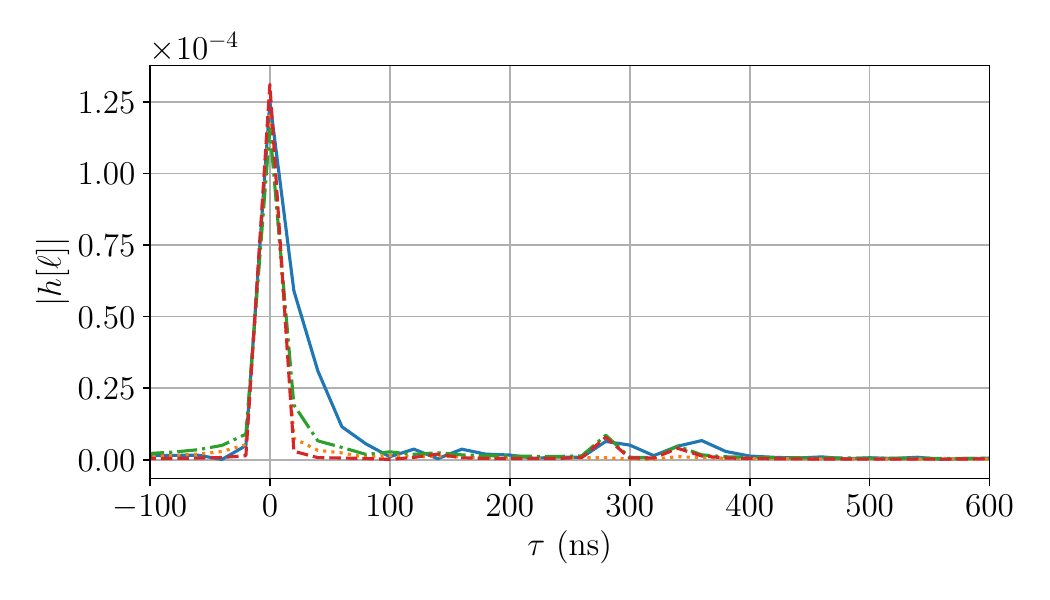}
  \end{subfigure}
  \hfill
  \begin{subfigure}{0.48\textwidth}
    \includegraphics[width=\linewidth]{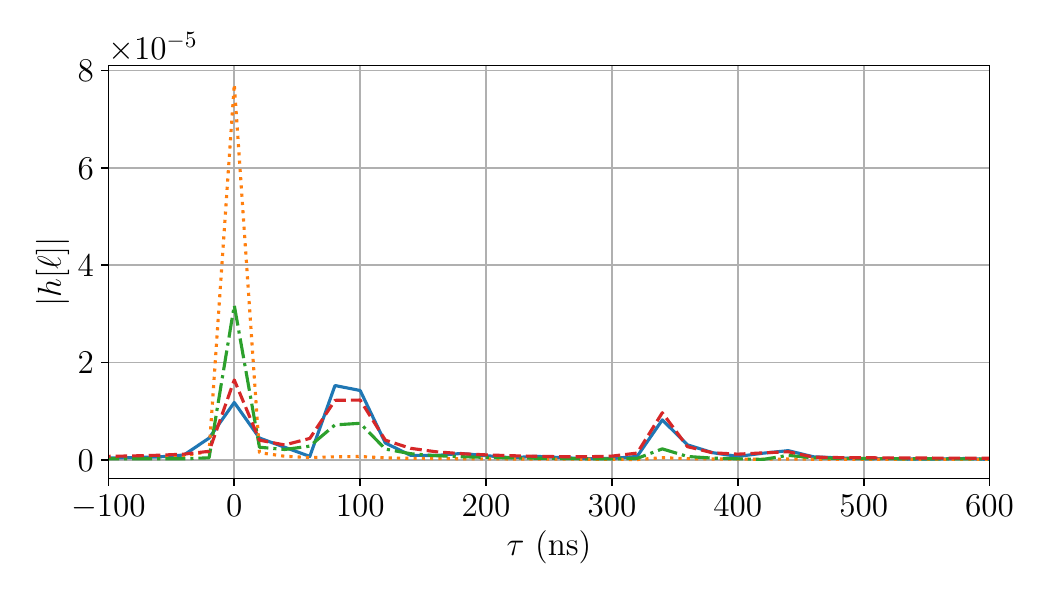}
  \end{subfigure}
  \caption{\gls{CIR} examples for different measurement positions and calibration techniques}
  \label{fig:cir}
\end{figure*}

Fig.~\ref{fig:cir} depicts measured and simulated \glspl{CIR} at four different positions. The ``ITU Materials'' generally fail to model paths with long propagation delays which can be partially compensated for through calibration using ``Learned Materials'' (cf. Fig.~\ref{fig:cdf_delay_spread}). However, due to the very coarse scene representation, this impacts the accuracy of early reflections. The fine-grained modeling capabilities of ``Neural Materials'' do not suffer from this problem and are capable of producing \glspl{CIR} that closely resemble the measurements. 

\begin{figure}
  \centering
   \includegraphics[width=\columnwidth]{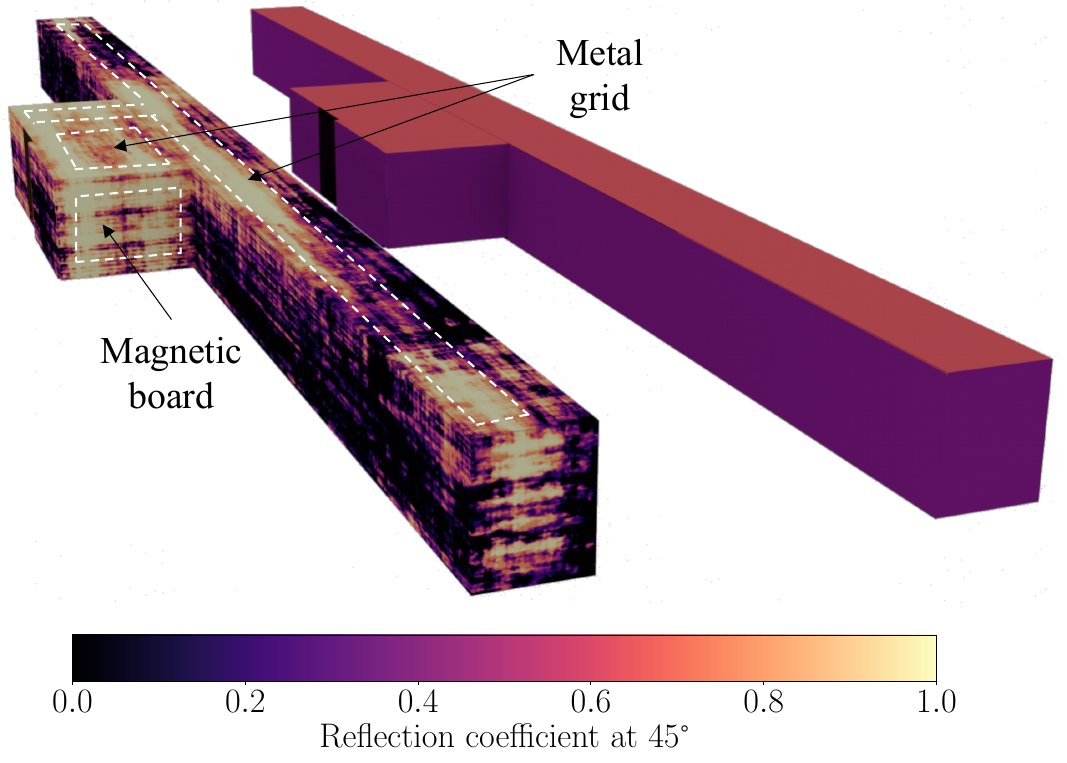}
  \caption{Visualization of the reflection coefficient across the scene: Neural Materials (left), Learned Materials (right).}
  \label{fig:reflectivity}
\end{figure}

Fig.~\ref{fig:reflectivity} illustrates the reflection coefficient ${r = \frac12(|r_\perp|^2 + |r_\parallel|^2)}$ at an incident angle of $\theta_i=\SI{45}{\degree}$; comparing ``Learned-'' and ``Neural Materials'' across the scene. 
The ``Neural Materials'' notably assign high values to the locations of the metallic grid on the ceiling and the magnetic board on a wall in the entrance hall, as seen in Fig.~\ref{fig:scenario}. While further study is needed, these observations suggest that ``Neural Materials'' can effectively represent objects' electromagnetic behaviors in a scene, offering a unique perspective on the environment through the lens of radio propagation.

\subsection{Discussion}
We have only scratched the surface of gradient-based calibration in this paper and many open questions remain to be answered. We will now briefly discuss a few of them:

\begin{itemize}
  \item The loss function \eqref{eq:loss} is non-convex in the scene parameters and there is no guarantee that gradient-based solutions will converge to the global minimum. It is not even clear if there is a unique optimal solution. For example, the same absolute value of a reflection coefficient \eqref{eq:fresnel-1} can be obtained for different values of conductivity and permittivity. Regularization of parameters during calibration may enforce plausible solutions. 

  \item Although all parameters have clear physical interpretations, the calibration does not necessarily lead to the same values which would be obtained by directly measuring them for each object. The reason for this is that calibration also compensates for errors in the geometry and measurements as well as modelling errors in the \gls{RT} process. For example, we have entirely ignored refraction in Section~\ref{sec:measurements}, which certainly plays an important role in our indoor scenario.

  \item The advantage of ``Neural Materials'' is their capability to represent material properties in a very fine-grained (continuous) manner across the entire scene. However, this comes at the cost of poor generalization performance to areas in which no measurements are available. Investigating their bias-variance trade-off is an important topic of future research. It is also interesting to study extensions or alternatives to the positional encoding \eqref{eq:pos_encoding}. 
  
  \item Our calibration method is agnostic to phase errors. One may learn location-specific phase correction terms through ``Neural Materials'' or investigate phase-error aware calibration techniques.
\end{itemize}

%% file: sections/conclusions.tex
\section{Conclusions}\label{sec:conclusions}
In this paper, we have introduced a novel gradient-based calibration method for differentiable ray tracers, enabling scene calibration akin to training a deep neural network. This method hinges on a carefully designed loss function, coupled with differentiable parameterizations of material properties, scattering behaviors, and antenna patterns. We have implemented our calibration method in the open-source ray tracer Sionna~RT \cite{sionna-rt} and validated it against indoor measurements from a distributed \gls{MIMO} channel sounder. This advancement in differentiable ray tracing, especially when combined with contemporary machine learning techniques, opens up exciting avenues for multidisciplinary research and holds immense potential for future explorations.